\documentclass[twoside]{article}
\usepackage{fleqn}
\usepackage{espcrc2}
\usepackage{graphicx}
\usepackage{epsfig}
\usepackage{amsmath}
\usepackage{cite}
\usepackage[figuresright]{rotating}

\newcommand{\SinEff}{$\sin^2\theta^{\mbox{\footnotesize lept}}_{\mbox{\footnotesize eff}}$}
\newcommand{\sineff}{\sin^2\theta^{\mbox{\footnotesize lept}}_{\mbox{\footnotesize eff}}}

\title{%
Two-loop Fermionic Electroweak Corrections \\
to the Effective Leptonic Weak Mixing Angle in the Standard Model
\thanks{Work supported in part 
  by TMR, European Community Human Potential Programme under contracts
  HPRN-CT-2002-00311 (EURIDICE), HPRN-CT-2000-00149 (Physics at
  Colliders), by Deutsche Forschungsgemeinschaft under contract SFB/TR
  9--03, and by the Polish State Committee for Scientific Research (KBN)
  under contract No. 2P03B01025.}}

\author{M. Awramik
\address{DESY, Platanenallee 6, D-15738 Zeuthen, Germany}
\address{Institute of Nuclear Physics PAS, Radzikowskiego 152,
  PL-31342 Cracow, Poland},
M. Czakon$^\text{ a}$
\address{Institute of Physics, University of Silesia, Uniwersytecka 4,
  PL-40007 Katowice, Poland},
A. Freitas
\address{Theoretical Physics Division, Fermilab, P. O. Box 500, Batavia, IL
  60510, USA},
G. Weiglein
\address{Institute for Particle Physics Phenomenology, University of
Durham, Durham DH1~3LE, UK}}

\begin{document}

\begin{abstract}
We give some details of the recently completed calculation of the
full two-loop fermionic corrections to the effective leptonic weak
mixing angle, \SinEff. Among others, we describe the C++ library
DiaGen/IdSolver, which was used to reduce the two-loop light fermion 
vertex diagrams to linear combinations of master integrals with
rational function coefficients.
\end{abstract}

\maketitle
%\maketitle\thispagestyle{empty}
%\pagestyle{empty}

%%%%%%%%%%%%%%%%%%%%%%%%%%%%%%%%%%%%%%%%%%%%%%%%%%%%%%%%%%%%%%%%%%%%%%%%%%%%%%%%

\section{INTRODUCTION}

Two masses play at present an extremely important r\^ole among the
parameters of the Standard Model. These are $M_W$ and $M_H$, the
masses of the W and Higgs bosons respectively. As long as experiments
will not reach a satisfactory level of precision for $M_W$ and will
not be able to give a value to $M_H$ by direct observation, we are
bound to seek indirect predictions for these observables based on
theoretical calculations of suitably chosen processes. It is well
known that $M_W$ can be predicted with the help of the precisely
measured muon decay lifetime, while $M_H$ from the Z peak observables,
most notably from the effective leptonic weak mixing angle, \SinEff.

The theoretical program for the W boson mass prediction has been
completed with an error estimate of 4 MeV \cite{mw}, which is even
below the expected precision of measurement at a future linear
collider, not to mention the much closer Large Hadron Collider (LHC),
where the error should be of the order of 15 MeV. This achievement
required among others the two-loop electroweak result finished with a
calculation of the bosonic part \cite{bosonic}, following a long
study of the fermionic part \cite{mt2,muon}. To be complete we have to
mention more than twenty years of various works starting from the
original analysis \cite{sirlin}, going through QCD corrections to the
one-loop result \cite{qcd2,qcd3} and finally some three-loop leading
contributions in the top quark mass \cite{faisst}.

In the case of the effective leptonic weak mixing angle, \SinEff, the
situation is not so satisfactory. Most of the corrections listed above
still apply, but a complete calculation at the two-loop level in the
electroweak interactions does not exist, and only the $m_t^4$ and $m_t^2$ terms in
the top quark mass expansion are available
\cite{Fleischer:1993ub,Degrassi:1996ps}. Since the current measurement has an
impressive precision, $\sineff = 0.23150 \pm 0.00016$ \cite{exp}, and
prospects are to reach an absolute error of $10^{-5}$ at a future
linear collider, it is still necessary to compute several
contributions. Recently, we took a step in this direction by
performing a complete calculation of the two-loop electroweak
contributions generated by diagrams with one or two closed fermion
loops \cite{sineff}. In this contribution, some aspects of this
calculation will be described. We will first present the method used
for the top quark vertex diagrams, then that for the light fermion
diagrams. Finally we will discuss some details of the C++ library
DiaGen/IdSolver \cite{idsolver} used for the decomposition of diagrams
to combinations of master integrals.

%%%%%%%%%%%%%%%%%%%%%%%%%%%%%%%%%%%%%%%%%%%%%%%%%%%%%%%%%%%%%%%%%%%%%%%%%%%%%%%%

\section{TOP QUARK CONTRIBUTIONS}

\SinEff is defined through the form factors of the leptonic Z boson
vertex. Namely, if the vertex is $i\; \overline{l} \gamma^\mu (g_V-g_A
\gamma_5)l \; Z_\mu$, then
\begin{equation}
  \label{definition}
  \sineff = \frac{1}{4}
  \left(1-\mbox{Re}\left(\frac{g_V}{g_A}\right)\right).
\end{equation}
The main complication in the calculation comes from the two-loop
vertices that one has to evaluate. All other irreducible two-loop
diagrams enter through renormalization constants and can be obtained
from propagator diagrams at nonvanishing external momentum at worst. The
algorithm to obtain numerical values of these has been completed in
\cite{Bauberger:1994by}.

We decided to divide the fermionic two-loop vertices in two groups,
one with top quark lines and one that has only light fermion
lines. The contribution of the former is expressed through three
massive parameters (we neglect the light fermion masses), $M_W$, $M_Z$
and $m_t$, and is thus a nontrivial function of two dimensionless
variables\footnote{The fermionic vertex diagrams do not depend on the
Higgs boson mass due to {\it CP} conservation.}. Although it is
conceivable that the result could be expressed in some closed analytic
form, this is absolutely not needed. It turns out that it is better to
exploit the smallness of the ratio $M_Z^2/m_t^2 \sim 1/4$, which
corresponds to decoupling of the top quark. Obviously the construction
of a high precision effective theory would be very difficult, since
high order tensor operators would be needed. Instead, we simply
performed a diagrammatic large mass expansion.

\begin{figure}
\begin{center}
\epsfig{file=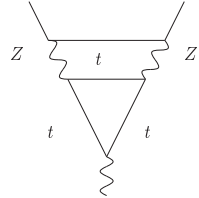,width=3.1cm}
\end{center}
\vspace{-.7cm}
\caption{\label{zz} Example of a diagram with a top quark subloop
  treated with the large top quark mass expansion.}
\end{figure}

We checked that inclusion of terms of order $(M_Z^2/m_t^2)^5$ is
sufficient to obtain an intrinsic precision of $10^{-7}$, which is by
far enough for practical purposes. An example of a scalar diagram
entering the calculation is given in Fig.~\ref{zz}. With
$x=M_Z^2/m_t^2$, the expansion reads
\begin{eqnarray}
&& \frac{x}{3}\zeta_{2} + \frac{x^2}{4}
	  \left(\frac{1}{3}\zeta_{2}-\frac{5}{9}+\frac{1}{3}\log x
	  \right) \nonumber \\
&&	  +\frac{x^3}{5} \left(
	  \frac{1}{9}\zeta_{2}-\frac{79}{240}+\frac{1}{4}\log x
	  \right)+\dots \;.
\end{eqnarray}
Numerically, this gives
\begin{equation}
0.1483\;-\; 0.0081\;-\;0.0019\;+\;0.0003\;+\dots \;.
\end{equation}
Such excellent convergence is typical of all neutral current
diagrams. The charged current diagrams do not converge so rapidly
presumably because of the splitting of the W boson lines into a
top-bottom pair as opposed to double top pairs in the previous case.

Let us stress that we performed the expansion exclusively for the
vertex diagrams. This means that the propagator diagrams from
renormalization, which contain one massive parameter more, the Higgs
boson mass, were evaluated by exact one-dimensional integral
representations. Our approach is therefore different from that of
\cite{Degrassi:1996ps}, where expansion was performed in $M_H$ as
well. In that case, though, only the leading and subleading terms in
$m_t$ were computed, whereas we obtained the complete result for the
fermionic two-loop contributions.

%%%%%%%%%%%%%%%%%%%%%%%%%%%%%%%%%%%%%%%%%%%%%%%%%%%%%%%%%%%%%%%%%%%%%%%%%%%%%%%%

\section{LIGHT FERMION CONTRIBUTIONS}

The light fermion vertex diagrams contain one scale less and are,
therefore, functions of only one dimensionless variable. It turned out,
that this case allows to perform an exact calculation and obtain the
result in  closed analytic form expressed through polylogarithms. To
this end we used the differential equation method \cite{Kotikov:hm}.

\begin{figure}
\begin{center}
\epsfig{file=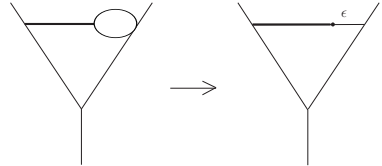,width=5.cm}
\end{center}
\vspace{-.7cm}
\caption{\label{subloop} Example of a light fermion subloop integrated using the
  differential equation technique. The thick line is massive.}
\end{figure}

An example of a scalar integral entering the calculation is given in
Fig.~\ref{subloop}. Since the subloop is massless it can be
integrated to give a massless line with a non-integer power of
the denominator. In fact using harmonic tensors it is also possible to
perform the full tensor reduction of any scalar integral corresponding
to this topology. To get rid of the higher powers of denominators we
devised a reduction scheme with the help of integration by parts
identities for general indices.

The integral corresponding to Fig.~\ref{subloop}, denoted by LF1,
satisfies the following differential equation
\begin{eqnarray} 
\label{diff}
&& \!\!\!\!\!\!\!\!\!\!\!\!\! M^2 \frac{d}{d M^2} \mbox{LF1}(M,m) = \\
  && \!\!\!\!\!\!\!\!\!\!\!\!\! \frac{1}{2}
  \frac{M^2}{M^2+m^2} ( (4-d)(4 +5 \frac{m^2}{M^2}) \; \mbox{LF1}(M,m)
  \nonumber \\ &&\!\!\!\!\!\!\!\!\!\!\!\!\! +  (10-3 d)\; \mbox{LF0}(M) - (2-d)\;
  \frac{1}{m^2}\mbox{T134}(0,0,m) ), \nonumber
\end{eqnarray}
where LF0 is the same integral without the massive line, T134
\cite{Bauberger:1994by} is the
sunset vacuum integral, $M^2$ is the external momentum squared and
$m$ is the mass on the massive line. Both LF1 and LF0 have been made
dimensionless with the help of the external momentum.

Equation~\ref{diff} may be integrated in the class of Nielsen
polylogarithms. The polylogarithmic terms of the finite part read
\begin{equation*}
  -\mbox{Li}_2 (-x)(-2 + 2\log (m^2) + 3\log (-x) + \log (1 + x))
\end{equation*}
\begin{equation}
  + 4\mbox{Li}_3 ( -x) - \mbox{S}_{1, 2} (-x) + \dots
\end{equation}
where $x=M^2/m^2$. We should note that LF1 has also been calculated in
\cite{kuhn}.

We were able to integrate analytically all of the integrals but
one. The last remaining integral was then evaluated
numerically. Recently, another work appeared \cite{bonciani}, where
several integrals of interest to us have also been
calculated. However, the authors of that calculation introduced an
extension of the polylogarithm class and represented their result in
terms of this extension. In the end, to obtain a numerical result one
would need to perform multiple integrals.

Finally, let us note that all integrals have been checked by different
expansions in physical and unphysical regimes and by numerical
integrations based on dispersion relations \cite{Bauberger:1994by}
and Feynman parameterizations \cite{Ghinculov:1994sd}.
We have also performed two independent calculations of the on-shell 
renormalization procedure and the necessary counterterms for establishing a
finite and meaningful result for \SinEff.

%%%%%%%%%%%%%%%%%%%%%%%%%%%%%%%%%%%%%%%%%%%%%%%%%%%%%%%%%%%%%%%%%%%%%%%%%%%%%%%%

\section{DIAGEN/IDSOLVER}

\begin{figure}
\begin{center}
\epsfig{file=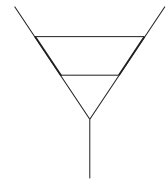,width=2.1cm}
\end{center}
\vspace{-.7cm}
\caption{\label{difficult} Most complicated light fermion prototype.}
\end{figure}

The calculation of the light fermion diagrams described in the last
section necessitates a reduction of integrals with irreducible
numerators and denominators with higher powers to a small set of
master integrals. In the case of the LF1 integral mentioned above, the
procedure is relatively simple, since after integration of the subloop
it is similar to a one-loop integral. However, in the case of
Fig.~\ref{difficult}, this is much more difficult.

To solve this and similar problems, a complete system has been
designed and programmed in C++. This system is built upon the DiaGen
\cite{idsolver} library which provides tools for diagram generation
and topological analysis. The new part, called IdSolver
\cite{idsolver}, implements, among others, the Laporta algorithm
\cite{laporta}. The major difference compared to the system described in
\cite{laporta}\footnote{see also \cite{anastasiou}.} is the automated
topological analysis part.

\begin{figure}
\begin{center}
\epsfig{file=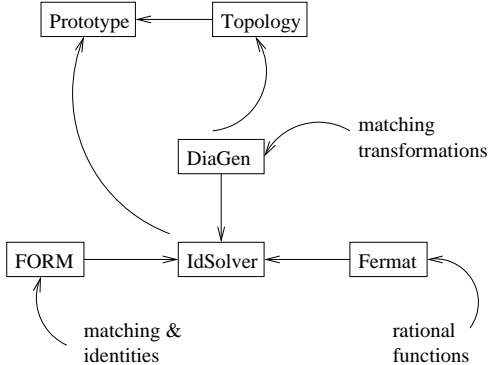,width=6.59cm}
\end{center}
\vspace{-.7cm}
\caption{\label{schema} Structure and interrelationships in IdSolver.}
\end{figure}

The general structure and interrelationships with other software used
by IdSolver are shown in Fig.~\ref{schema}. Internally, the library
is organized around two pairs of classes

\begin{itemize}
  \item {\bf Prototype and PrototypeList}
    
    A prototype is a topology with masses assigned to lines (a mass of
    an external line is defined through its momentum), {\it i.e.}\ a
    colored undirected graph in graph theoretical terminology.
    Identification of isomorphic prototypes is performed either
    through momentum patterns for vacuum integrals or through
    topological isomorphism modulo tadpoles. The latter means that the
    original prototype is decomposed into its subgraph without
    tadpoles and tadpoles separately, then isomorphism tests are
    performed on each component. The reason for such sophistication is
    that (topologically) different prototypes may be associated with
    the same integral. The procedure above is unique only for vacuum
    graphs. However, in all problems considered in
    practice\footnote{this encompasses QED vertices and boxes
    \cite{bhabha}, electroweak vertices and tadpoles up to four loops.}
    no duplicate integrals have been generated. Starting from box
    graphs, there is a new problem associated with the exchanges of
    external lines that have the same momentum squared. An algorithm
    has been implemented which identifies two prototypes as isomorphic
    only if the necessary interchange does not permute the invariants
    ({\it e.g.}\ s, t and u for boxes). Besides prototype isomorphism,
    the Prototype class provides, among others, the topological
    symmetry group, automatically assigned momenta (with optimization
    of distribution) and numerator/denominator cancellation rules.
    
    The PrototypeList class manages hierarchies of prototypes. Upon
    insertion of a prototype it generates all of its subprototypes
    obtained by canceling lines together with matching rules, {\it
    i.e.}\ momentum shifting transformations between the prototype and
    its subprototypes. Simultaneously, it writes administration files
    to disk starting from declarations needed by 
    FORM \cite{Vermaseren:2000nd} and ending with
    precomputed Integration By Parts (IBP), Lorentz Invariance (if
    requested by user) or other identities (user
    defined). Subsequently it manages the solution of the system
    starting from the simplest prototypes (those that have the least
    number of lines).

  \item {\bf Integral and IntegralList}

    From the point of view of the system, an integral is just a name
    and list of indices, which represent powers of the irreducible
    numerators and denominators. Integrals may be equivalent, if
    they do not have numerators and are equal by symmetry. This
    property is automatically assigned by the system based on the
    symmetry group of the prototype. Once solved, integrals have
    associated expressions, which are linear combinations of other
    integrals (masters) with rational function coefficients.

    The system of integrals is represented by the IntegralList
    class. The latter provides a method for updating the list with an
    identity generated by FORM and for writing the solution to
    disk. It also allows for master integral identification and keeps
    various statistical information.
\end{itemize}

\vspace{-0.035cm}

The library has several other classes, the description of
which we will skip.

The rational function arithmetic needed may be done by any external software
by means of a specialized thin interface class. In practice,
Maple, Mathematica and Fermat \cite{fermat} have been used. The latter
has proved to be the fastest and most economic in terms of resources required.

An important step in the calculation is the master identification. To
be safe, one should solve the system to high powers of numerators and
denominators. If the expressions were kept exact, this would be time
consuming and unnecessary. To solve this problem, evaluation
homomorphisms are used, {\it i.e.}\ the system is solved by projecting
the coefficients to the rational numbers field with suitably chosen
values of the parameters.

The library DiaGen/IdSolver is a powerful tool many
possible applications. It is being currently used to calculate among
others the two-loop bosonic corrections to \SinEff.

%%%%%%%%%%%%%%%%%%%%%%%%%%%%%%%%%%%%%%%%%%%%%%%%%%%%%%%%%%%%%%%%%%%%%%%%%%%%%%%%


\begin{thebibliography}{00}

\bibitem{mw}
M.~Awramik, M.~Czakon, A.~Freitas and G.~Weiglein,
%``Precise prediction for the W-boson mass in the standard model,''
Phys.\ Rev.\ D {\bf 69} (2004) 053006.
%%CITATION = HEP-PH 0311148;%%

\bibitem{bosonic}
M.~Awramik and M.~Czakon,
%``Complete two loop bosonic contributions to the muon lifetime in the
%standard model,''
Phys.\ Rev.\ Lett.\  {\bf 89} (2002) 241801,
%%CITATION = HEP-PH 0208113;%%
%``Two loop electroweak bosonic corrections to the muon decay lifetime,''
Nucl.\ Phys.\ Proc.\ Suppl.\  {\bf 116} (2003) 238;
A.~Onishchenko and O.~Veretin,
%``Two-loop bosonic electroweak corrections to the muon lifetime and
%M(Z) M(W) interdependence,''
Phys.\ Lett.\ B {\bf 551} (2003) 111;
%%CITATION = HEP-PH 0209010;%%
M.~Awramik, M.~Czakon, A.~Onishchenko and O.~Veretin,
%``Bosonic corrections to Delta(r) at the two loop level,''
Phys.\ Rev.\ D {\bf 68} (2003) 053004.
%%CITATION = HEP-PH 0209084;%%

\bibitem{Fleischer:1993ub}
R.~Barbieri, M.~Beccaria, P.~Ciafaloni, G.~Curci and A.~Vicere,
%``Radiative correction effects of a very heavy top,''
Phys.\ Lett.\ B {\bf 288} (1992) 95
[Erratum-ibid.\ B {\bf 312} (1993) 511],
%[hep-ph/9205238].
%%CITATION = HEP-PH 9205238;%%
%R.~Barbieri, M.~Beccaria, P.~Ciafaloni, G.~Curci and A.~Vicere,
%``Two loop heavy top effects in the Standard Model,''
Nucl.\ Phys.\ B {\bf 409} (1993) 105;
%%CITATION = NUPHA,B409,105;%%
J.~Fleischer, O.~V.~Tarasov and F.~Jegerlehner,
%``Two loop heavy top corrections to the rho parameter: A Simple formula valid
%for arbitrary Higgs mass,''
Phys.\ Lett.\ B {\bf 319} (1993) 249,
%%CITATION = PHLTA,B319,249;%%
%J.~Fleischer, O.~V.~Tarasov and F.~Jegerlehner,
%``Two loop large top mass corrections to electroweak parameters: Analytic
%results valid for arbitrary Higgs mass,''
Phys.\ Rev.\ D {\bf 51} (1995) 3820.
%%CITATION = PHRVA,D51,3820;%%

\bibitem{mt2}
G.~Degrassi, P.~Gambino and A.~Vicini,
%``Two-loop heavy top effects on the MZ-MW interdependence,''
Phys.\ Lett.\ B {\bf 383} (1996) 219;
%%CITATION = HEP-PH 9603374;%%
G.~Degrassi, P.~Gambino and A.~Sirlin,
 %``Precise calculation of M(W), sin**2(Theta(W)(M(Z))), and
%sin**2(Theta(eff)(lept)),''
Phys.\ Lett.\ B {\bf 394}, 188 (1997).
%%CITATION = HEP-PH 9611363;%%

\bibitem{muon}
A.~Freitas, W.~Hollik, W.~Walter and G.~Weiglein,
%``Complete fermionic two-loop results for the M(W)-M(Z)
%interdependence,''
Phys.\ Lett.\ B {\bf 495} (2000) 338
[Erratum-ibid.\ B {\bf 570} (2003) 260],
%%CITATION = HEP-PH 0007091;%%
%A.~Freitas, W.~Hollik, W.~Walter and G.~Weiglein,
%``Electroweak two-loop corrections to the M(W) - M(Z) mass correlation
%in  the standard model,''
Nucl.\ Phys.\ B {\bf 632} (2002) 189
[Erratum-ibid.\ B {\bf 666} (2003) 305]; 
%%CITATION = HEP-PH 0202131;%%
M.~Awramik and M.~Czakon,
%``Complete two loop electroweak contributions to the muon lifetime in
%the  standard model,''
Phys.\ Lett.\ B {\bf 568} (2003) 48.
%%CITATION = HEP-PH 0305248;%%

\bibitem{sirlin}
A.~Sirlin,
%``Radiative Corrections In The SU(2)-L X U(1) Theory: A Simple
%Renormalization Framework,''
Phys.\ Rev.\ D {\bf 22} (1980) 971; 
%%CITATION = PHRVA,D22,971;%%
W.~J.~Marciano and A.~Sirlin,
%``Radiative Corrections To Neutrino Induced Neutral Current Phenomena
%In The SU(2)-L X U(1) Theory,''
Phys.\ Rev.\ D {\bf 22} (1980) 2695
[Erratum-ibid.\ D {\bf 31} (1985) 213].
%%CITATION = PHRVA,D22,2695;%%

\bibitem{qcd2}
A.~Djouadi and C.~Verzegnassi,
%``Virtual Very Heavy Top Effects In Lep / Slc Precision Measurements,''
Phys.\ Lett.\ B {\bf 195} (1987) 265; 
%%CITATION = PHLTA,B195,265;%%
A.~Djouadi,
%``O (Alpha Alpha-S) Vacuum Polarization Functions Of The Standard Model
%Gauge Bosons,''
Nuovo Cim.\ A {\bf 100} (1988) 357; 
%%CITATION = NUCIA,A100,357;%%
B.~A.~Kniehl,
%``Two Loop Corrections To The Vacuum Polarizations In Perturbative
%QCD,''
Nucl.\ Phys.\ B {\bf 347} (1990) 86; 
%%CITATION = NUPHA,B347,86;%%
F.~Halzen and B.~A.~Kniehl,
%``Delta r beyond one loop,''
Nucl.\ Phys.\ B {\bf 353} (1991) 567; 
%%CITATION = NUPHA,B353,567;%%
B.~A.~Kniehl and A.~Sirlin,
%``Dispersion relations for vacuum polarization functions in electroweak
%physics,''
Nucl.\ Phys.\ B {\bf 371} (1992) 141; 
%%CITATION = NUPHA,B371,141;%%
B.~A.~Kniehl and A.~Sirlin,
%``On the effect of the t anti-t threshold on electroweak parameters,''
Phys.\ Rev.\ D {\bf 47} (1993) 883; 
%%CITATION = PHRVA,D47,883;%%
A.~Djouadi and P.~Gambino,
%``Electroweak gauge bosons self-energies: Complete QCD corrections,''
Phys.\ Rev.\ D {\bf 49} (1994) 3499
[Erratum-ibid.\ D {\bf 53} (1994) 4111].
%%CITATION = HEP-PH 9309298;%%

\bibitem{qcd3}
L.~Avdeev, J.~Fleischer, S.~Mikhailov and O.~Tarasov,
%``0 (alpha alpha-s**2) correction to the electroweak rho parameter,''
Phys.\ Lett.\ B {\bf 336} (1994) 560
[Erratum-ibid.\ B {\bf 349} (1994) 597]; 
%%CITATION = HEP-PH 9406363;%%
K.~G.~Chetyrkin, J.~H.~K\"uhn and M.~Steinhauser,
%``Corrections of order O (G(F) M(t)**2 alpha-s**2) to the rho
%parameter,''
Phys.\ Lett.\ B {\bf 351} (1995) 331; 
%%CITATION = HEP-PH 9502291;%%
K.~G.~Chetyrkin, J.~H.~K\"uhn and M.~Steinhauser,
%``QCD corrections from top quark to relations between electroweak
%parameters to order alpha-s**2,''
Phys.\ Rev.\ Lett.\  {\bf 75} (1995) 3394.
%%CITATION = HEP-PH 9504413;%%

\bibitem{faisst}
M.~Faisst, J.~H.~K\"uhn, T.~Seidensticker and O.~Veretin,
%``Three loop top quark contributions to the rho parameter,''
Nucl.\ Phys.\ B {\bf 665} (2003) 649.
%%CITATION = HEP-PH 0302275;%%

\bibitem{Degrassi:1996ps}
G.~Degrassi, P.~Gambino and A.~Sirlin,
 %``Precise calculation of M(W), sin**2(Theta(W)(M(Z))), and
%sin**2(Theta(eff)(lept)),''
Phys.\ Lett.\ B {\bf 394} (1997) 188.
%%CITATION = HEP-PH 9611363;%%

\bibitem{exp}
The LEP EWWG and LEP Collaborations,
 %``A combination of preliminary electroweak measurements and 
%constraints on the
%standard model,''
hep-ex/0312023.
%%CITATION = HEP-EX 0312023;%%

\bibitem{sineff}
M.~Awramik, M.~Czakon, A.~Freitas and G.~Weiglein, hep-ph/0407317.

\bibitem{idsolver} M.~Czakon, DiaGen/IdSolver ({\it unpublished}).

\bibitem{Bauberger:1994by}
G.~Weiglein, R.~Scharf and M.~B\"ohm,
%``Reduction of general two loop selfenergies to standard scalar integrals,''
Nucl.\ Phys.\ B {\bf 416} (1994) 606;
%%CITATION = HEP-PH 9310358;%%
S.~Bauberger, F.~A.~Berends, M.~B\"ohm and M.~Buza,
%``Analytical and numerical methods for massive two loop selfenergy diagrams,''
Nucl.\ Phys.\ B {\bf 434} (1995) 383.
%%CITATION = HEP-PH 9409388;%%

\bibitem{Kotikov:hm}
A.~V.~Kotikov,
 %``Differential Equations Method: The Calculation Of Vertex Type Feynman
%Diagrams,''
Phys.\ Lett.\ B {\bf 259} (1991) 314;
%%CITATION = PHLTA,B259,314;%%
E.~Remiddi,
%``Differential equations for Feynman graph amplitudes,''
Nuovo Cim.\ A {\bf 110} (1997) 1435.
%%CITATION = HEP-TH 9711188;%%

\bibitem{kuhn}
B.~Feucht, J.~H.~K\"uhn and S.~Moch,
 %``Fermionic and scalar corrections for the Abelian form factor at two
%loops,''
Phys.\ Lett.\ B {\bf 561} (2003) 111.
%%CITATION = HEP-PH 0303016;%%

\bibitem{bonciani}
U.~Aglietti and R.~Bonciani,
 %``Master integrals with one massive propagator for the two-loop  electroweak
%form factor,''
Nucl.\ Phys.\ B {\bf 668} (2003) 3; 
%%CITATION = HEP-PH 0304028;%%
U.~Aglietti and R.~Bonciani,
 %``Master integrals with 2 and 3 massive propagators for the 2-loop electroweak
%form factor: Planar case,''
hep-ph/0401193.
%%CITATION = HEP-PH 0401193;%%

\bibitem{Ghinculov:1994sd}
A.~Ghinculov and J.~J.~van der Bij,
%``Massive two loop diagrams: The Higgs propagator,''
Nucl.\ Phys.\ B {\bf 436} (1995) 30.
%%CITATION = HEP-PH 9405418;%%

\bibitem{laporta}
S. Laporta and E. Remiddi, Phys. Lett. B379 (1996) 283.
S. Laporta, Int. J. Mod. Phys. A15 (2000) 5087.

\bibitem{anastasiou}
Y.~Schr\"oder,
%``Automatic reduction of four-loop bubbles,''
Nucl.\ Phys.\ Proc.\ Suppl.\  {\bf 116} (2003) 402; 
%%CITATION = HEP-PH 0211288;%%
C.~Anastasiou and A.~Lazopoulos,
%``Automatic integral reduction for higher order perturbative calculations,''
hep-ph/0404258.
%%CITATION = HEP-PH 0404258;%%

\bibitem{bhabha}
M.~Czakon, J.~Gluza and T.~Riemann,
 %``A complete set of scalar master integrals for massive 2-loop Bhabha
%scattering: Where we are,''
hep-ph/0406203.
%%CITATION = HEP-PH 0406203;%%

\bibitem{Vermaseren:2000nd}
J.~A.~Vermaseren,
math-ph/0010025.

\bibitem{fermat}
R. H. Lewis, Fermat,
% http://www.bway.net/~lewis/.
www.bway.net/\verb+~+lewis/.

\end{thebibliography}
\end{document}